# *Clathrate formation in the systems Sr-Cu-Ge and {Ba,Sr}-Cu-Ge*

**I. Zeiringer[1], A. Grytsiv[1], F. Kneidinger[2], E. Royanian[2], E. Bauer[2], G. Giester[3], M. Falmbigl[1], P. Rogl[1,*]**

[1]*Institute of Physical Chemistry, University of Vienna, Währingerstrasse 42, A-1090 Wien, Austria*
[2]*Institute of Solid State Physics, Vienna University of Technology, Wiedner Hauptstrasse 8-10, A-1040 Wien, Austria*
[3]*Institute of Mineralogy and Crystallography, University of Vienna, Althanstrasse14, A-1090 Wien, Austria*

## ABSTRACT

In the ternary system Sr-Cu-Ge, a clathrate type-I phase, $Sr_8Cu_{5.3}Ge_{40.7}$ (a = 1.06311(3), exists close to the Zintl limit in a small temperature interval. $Sr_8Cu_{5.3}Ge_{40.7}$ decomposes eutectoidally on cooling at 730°C into (Ge), $SrGe_2$ and $\tau_1$-$SrCu_{2-x}Ge_{2+x}$. Phase equilibria at 700°C have been established for the Ge rich part and are characterized by the appearance of only one ternary compound, $\tau_1$-$SrCu_{2-x}Ge_{2+x}$, which crystallizes with the $ThCr_2Si_2$ structure type and forms a homogeneity range up to x=0.4 (a = 0.42850(4), c = 1.0370(1) nm).

Additionally, the extent of the clathrate type-I solid solution $Ba_{8-x}Sr_xCu_yGe_{46-y}$ ($5.2 \leq y \leq 5.4$) has been studied at various temperatures. The clathrate type-I crystal structure (space group $Pm\bar{3}n$) has been proven by X-ray single crystal diffraction on two single crystals with composition $Sr_8Cu_{5.3}Ge_{40.7}$ (a = 1.06368(2) nm) and $Ba_{4.9}Sr_{3.1}Cu_{5.3}Ge_{40.7}$ (a = 1.06748(2) nm) measured at 300, 200 and 100 K. From the temperature dependency of the lattice parameters and the atomic displacement parameters, the thermal expansion coefficients, the Debye- and Einstein-temperatures and the speed of sound have been determined.

From heat capacity measurements of $Sr_8Cu_{5.3}Ge_{40.7}$ at low temperatures, the Sommerfeld coefficient and the Debye temperature have been extracted, whereas from a detailed analysis of these data at higher temperatures, Einstein branches of the phonon dispersion relation have been derived and compared with those obtained from the atomic displacement parameters.

Electrical resistivity measurements of $Sr_8Cu_{5.3}Ge_{40.7}$ reveal a rather metallic behaviour in the low temperature range (< 300 K).

* Corresponding author: Tel.: +43-1-4277-52456; fax: +43-1-4277-9524
*E-mail address:* peter.franz.rogl@univie.ac.at (P. Rogl)

# 1. INTRODUCTION

In various cases, the clathrate type-I solid solution offers the possibility to tune its promising thermoelectric properties from a metal towards an insulator and to obtain *n*- and *p*-type properties by varying the transition metal content in the crystal structure [1-4]. In the primitive cubic clathrate type-I, the transition metal atoms and atoms from the 3$^{rd}$ or 4$^{th}$ main group of the periodic table, form a 3-dimensional network of cages, which are filled by electropositive elements like Ca, Sr, Ba, Eu or Yb. In contrast to the large number of known intermetallic clathrates of type-I with barium as an electropositive guest atom inside the cages, the number of representatives reported with strontium is rather limited. Eisenmann et al. [5] were the first to report on the formation and crystal structure of the compounds $Sr_8Al_{16}Si_{30}$, $Sr_8Ga_{16}Si_{30}$ and $Sr_8Ga_{16}Ge_{30}$. Besides these examples, the only further Sr-based ternary clathrate type-I compound, which has been found so far, is $Sr_8Zn_8Ge_{38}$ [6]. Additionally, several investigations have been published dealing with the substitution of a forth element (a) either in the framework (e.g., $Sr_8Ga_{16}Ge_{30-x}Sn_x$ ($0 \leq x \leq 12$) [7] and $Sr_8Al_xGa_{16-x}Si_{30}$ (clathrate type-I, $0 \leq x \leq 7$ and type-VIII, $8 \leq x \leq 13$) [8]) or (b) as a second guest inside the cages (f. e. $Ba_{8-x}Sr_xAu_6Ge_{46-x}$ ($0 \leq x \leq 4$) [9] and $Ba_{8-x}Sr_xAl_{14}Si_{32}$ ($0.6 \leq x \leq 1.3$) [10]).

The present work will focus on (I) the formation and crystal structure of type-I clathrates in the systems Sr-Cu-Ge(Si,Sn) and Ba-Sr-Cu-Ge; (II) on phase relations in the Ge-rich part of the Sr-Cu-Ge system at 700°C and (III) on low temperature physical properties (electrical resistivity and specific heat) of $Sr_8Cu_{5.3}Ge_{40.7}$. The investigations will rely on single crystal X-ray diffraction at different temperatures (300, 200, 100 K), X-ray powder diffraction, differential thermal analysis and electron probe micro analysis,

# 2. EXPERIMANTAL DETAILS

All samples of about 0.2-1g were prepared by arc melting mixtures of the pure metal ingots (Sr 99.9, Ba 99.9, Cu 99.9, Ge 99.9999, Si 99.9999, Sn 99.9 mass%, Alfa Aesar) under an argon atmosphere. The molten buttons were turned over and remelted at least 3 times to ensure homogeneity with weight losses smaller than 1 mass% in all cases. Afterwards, the samples were cut into several pieces. One of them was used for characterization in as cast state, the others were vacuum-sealed in quartz tubes and annealed at various temperatures between 600 and 800°C.

Details of the various techniques of characterization such as scanning electron microscopy (SEM, including electron probe microanalysis (EPMA) on a Zeiss Supra 55 VP operated at 20 kV using EDX detection for the quantitative analysis), differential thermal analysis (DTA) and crystal structure analysis (X-ray powder diffraction (XPD) and X-ray single crystal diffraction (XSCD)) have been described in our previous papers [11, 12].

Electrical resistivity has been measured between 2 and 300 K with a standard four probe technique in a $^4$He-bath cryostat. The relaxation time method was used to measure specific heat down to 2 K in a conventional Quantum Design PPMS.

## 3. RESULTS AND DISCUSSION

### *3.1 Clathrates type-I in the systems Sr-Cu-Ge(Si,Sn) and Ba-Sr-Cu-Ge*

In the ternary system Sr-Cu-Ge, a clathrate type-I phase ($\kappa_I$) was found to exist in only a very small temperature interval with an almost point composition close to the Zintl limit $Sr_8Cu_{5.3}Ge_{40.7}$ (a = 1.06311(3), Ge standard). The $\kappa_I$-phase decomposes eutectoidally already at 730°C (from DTA measurements) according to the isothermal reaction: $\kappa_I \Leftrightarrow$ (Ge) + $SrGe_2$ + $SrCu_{2-x}Ge_{2+x}$ ($\tau_1$, for crystallographic details on the phases see Table I). Figure 1a shows this decomposition, starting at the grain boundaries of a sample annealed at 600°C; Fig. 1b is a close-up with a higher magnification. Although it was possible to receive a large amount of the clathrate phase in small (and thus rapidly quenched) arc-melted buttons (< 0.25 g), the decomposition of the clathrate phase is already visible at the top of the samples due to slower cooling (see figure 1c). All other attempts to produce larger single-phase samples of the ternary clathrate by quenching from high temperatures, including from liquid state, failed. Due to the presence of $SrGe_2$, multiphase alloys decompose rather quickly on air.

We further investigated the substitution of $\kappa_I$ via a forth element as a possible stabilizer either as a guest or a framework atom, as it was observed for the binary clathrate type-I $Ba_8Ge_{43}\square_3$, where the substitution with transition metals like silver leads to a strong stabilization [4, 12]. Concerning the framework, a partial substitution of Ge by Si or Sn proved unsuccessful by analyzing alloys with nominal composition $Sr_8Cu_5Si_xGe_{46-x}$ (x = 5, 10, 20, 30) and $Sr_8Cu_5Sn_2Ge_{44}$ in as cast state and after annealing at 750°C. Figure 1d shows the microstructure of the alloy $Sr_8Cu_5Si_5Ge_{41}$ in as cast state. The main phase is (Ge,Si), with a rather inhomogeneous distribution of Si. A small amount of silicon is also solved in the ternary compound $\tau_2$. The alloy $Sr_8Cu_5Sn_2Ge_{44}$ in as cast state (see figure 1e) also contains 3 phases, namely (Ge), $SrGe_2$ and a ternary phase $\tau_3$. A small amount of tin solves in (Ge) and $\tau_3$. Crystallographic details and EPMA data for these alloys are compiled in Table II.

Replacement of Sr by Ba, however, leads to the formation of a quaternary solution of clathrate type-I $Ba_{8-x}Sr_xCu_yGe_{46-y}$ ($5.2 \leq y \leq 5.4$), which is confined to the Ba-rich side and does not cover the region near $Sr_8Cu_{5.3}Ge_{40.7}$. Our present investigations of the clathrate type-I solid solution $Ba_{8-x}Sr_xCu_yGe_{46-y}$ are focused on the region with approximately five copper atoms (y = 5.2 - 5.4) per formula unit and strontium contents x between 0.5 and 8. In as cast state, all prepared samples showed a mixture of two clathrate type-I phases, namely $Sr_8Cu_{5.3}Ge_{40.7}$ ($\kappa_I$) and $Ba_{8-x}Sr_xCu_yGe_{46-y}$ ($\kappa_{I\text{-}Sr,Ba}$) with clearly defined phase boundaries in between. Figure 1f shows the microstructure of an alloy with nominal composition $Ba_{0.5}Sr_{7.5}Cu_{5.3}Ge_{40.7}$ in as cast state, indicating primary crystallization of $\kappa_{I\text{-}Sr,Ba}$ with composition $Ba_{2.4}Sr_{5.6}Cu_{5.3}Ge_{40.7}$ (from EPMA) followed by the crystallization of small amounts of $SrGe_2$ and $\kappa_I$ with composition $Sr_8Cu_{5.2}Ge_{40.8}$. After annealing at 760°C (figure 1g), large grains of both clathrate phases can be seen ($\kappa_{I\text{-}Sr,Ba}$ with composition $Ba_{4.8}Sr_{3.2}Cu_{5.2}Ge_{40.8}$ and $\kappa_I$), whereas after annealing at 700°C, $\kappa_I$ already almost decomposes and mainly large crystallites of $\kappa_{I\text{-}Sr,Ba}$ remain with different Sr/Ba ratios from the center ($Ba_{3.9}Sr_{4.1}Cu_{5.4}Ge_{40.6}$) to the edge ($Ba_{2.7}Sr_{5.3}Cu_{5.4}Ge_{40.6}$) of the grain (see figure 1h). The homogeneity region of the clathrate solid solution $Ba_{8-x}Sr_xCu_{\sim 5.3}Ge_{40.7}$ extends from x = 0 to x ~ 5.6 in as cast state. No measurable Ba content has been found in the Sr-clathrate $\kappa_I$. The dependency of the lattice parameter with increasing Sr content from x = 0 to x = 8 is shown in figure 2. The exact composition has been determined with EPMA. As can be seen, the lattice parameters almost linearly decrease with increasing Sr-content and the observed deviations are probably due to the slightly different Cu-content. Detailed information on the crystal structure of the clathrates type-I $Sr_8Cu_{5.3}Ge_{40.7}$ and $Ba_{4.9}Sr_{3.1}Cu_{5.3}Ge_{40.7}$ is given in the next chapter. Table II summarizes the results from EPMA and XPD for the alloys in figure 1 and they are correspondingly labeled from a) to h).

*3.2 Crystal structure of the clathrates type-I $Sr_8Cu_{5.3}Ge_{40.7}$ and $Ba_{4.9}Sr_{3.1}Cu_{5.3}Ge_{40.7}$*

Crystal specimens suitable for X-ray single crystal diffraction (performed at 300, 200 and 100 K) were isolated from samples with nominal composition $Sr_8Cu_{5.3}Ge_{40.7}$ in as cast state and $Ba_4Sr_4Cu_{5.3}Ge_{40.7}$ annealed at 800°C. For the crystal structure solution and refinement, the program Oscail [13, 14] has been used and the program Structure Tidy [15] was applied for subsequent standardization. The crystal structure was successfully solved for both compounds with direct methods in the cubic primitive space group $Pm\bar{3}n$, as typical for the clathrate type-I crystal structure.

The refinements for both crystals unambiguously showed a mixed occupancy of Cu and Ge in 6d whereas the Wyckoff sites 16i and 24k are fully occupied by Ge. In case of $Ba_{4.9}Sr_{3.1}Cu_{5.3}Ge_{40.7}$, the refinement furthermore revealed a mixed occupancy of Ba and Sr in the crystallographic sites 2a and 6c (atoms AE1 and AE2). No vacancies were detected in the crystal structures. Final refinements for both crystals were carried out introducing anisotropic atom displacement parameters (ADP's) and converged to R-values lower than $R_F<0.022$ and residual electron densities below 1.5 $e^-/Å^3$. The chemical formulae derived from these X-ray refinements ($Ba_{4.86}Sr_{3.14}Cu_{5.36}Ge_{40.64}$ and $Sr_8Cu_{5.36}Ge_{40.64}$) are in good agreement with the composition obtained from EPMA ($Ba_{4.2}Sr_{3.8}Cu_{5.3}Ge_{40.8}$ and $Sr_8Cu_{5.3}Ge_{40.7}$). Details of the structure refinement at 300 K can be found in Table III and of the structure refinement at 200 and 100 K in the supplementary information. X-ray powder diffraction intensities collected from polycrystalline samples are in good agreement with the intensities calculated from the structural models obtained from the single crystal studies.

The atom site distribution among Ba,Sr in $Ba_{4.9}Sr_{3.1}Cu_{5.3}Ge_{40.7}$, clearly shows that Ba,Sr atoms are sharing both centers of the cages, however, the smaller Sr-atoms reveal preference for the 2a-site in the smaller pentagondocehadral cage (occ.=0.805) with respect to a much lower Sr-occupancy of the 6c site (standardized setting) at the center of the larger tetrakaidekahedral cage (occ.=0.255). A mixed occupancy of both sites has also been found for $Eu_4Sr_4Ga_{16}Ge_{30}$ by synchrotron X-ray powder diffraction studies [16], whereas full atom order with the smaller atom in the pentagon dodecahedron has been reported for $K_6Eu_2Ga_{10}Ge_{36}$ [17], $K_6Eu_2Zn_5Ge_{41}$ [17], $K_6Eu_2Cd_5Ge_{41}$ [17] and $Eu_2Ba_6M_xSi_{46-x}$ (M=Cu, Al, Ga) [18] from X-ray single crystal data.

The anisotropic displacement parameters (ADP) of the guest atoms inside the large tetrakaidecahedron are usually much larger than those of the guests inside the smaller pentagondodecahedron (see Table IV) but for $Sr_8Cu_{5.3}Ge_{40.7}$, the ADP's of Sr2-atoms ($U_{11}$=0.0422(7); $U_{22}$=$U_{33}$=0.0802(7) Å²) were about 2 times larger applying the same structure model. Figure 3 shows a plot of the electron density of the atoms in the 6c site at 100 K (difference Fourier synthesis) illustrating a rather unitary round shape in case of $Ba_{4.9}Sr_{3.1}Cu_{5.3}Ge_{40.7}$ (Fig. 3b), whereas hints are visible for an off-center position in $Sr_8Cu_{5.3}Ge_{40.7}$ (Fig. 3a). This off-centering could be realized by placing the strontium atoms (Sr2-atoms) in a 24 k site with an occupation of ~1/4. Off-centering has already been successfully applied to other ternary type-I clathrates with Sr as guest ions like $Sr_8Zn_8Ge_{38}$ or $Sr_8Ga_{16}Ge_{30}$ [6]. For $Sr_8Cu_{5.3}Ge_{40.7}$ the distances between the off-center Sr2 atoms at 300 K are 2x Sr2-Sr2 0.0398 nm and 1x Sr2-Sr2 0.0486 nm and decrease only slightly with decreasing temperature (100 K: 2x Sr2-Sr2 0.0371 nm and 1x Sr2-Sr2 0.0421 nm); they are in good agreement with those in $Sr_8Zn_8Ge_{38}$. The results of the crystal structure refinement at 300 K are summarized in Table III (results of the structure refinement at lower temperatures can be found in the supplementary information).

Least squares fits as a function of temperature of the ADP's (see Fig.4) for AE2 or Sr2 (off-center model) for both crystals with the Einstein model (Eq.(1)) and with the Debye model (Eq.(2)) for the framework atoms

$$U_{ii} = \frac{\hbar^2}{2mk_B\Theta_{E,ii}}\coth\frac{\Theta_{E,ii}}{2T} + d^2 \qquad (1)$$

$$U_{iso} = \frac{3\hbar^2 T}{mk_B\Theta_D^2}\left[\frac{T}{\Theta_D}\int_0^{\Theta_D/T}\frac{x}{e^x-1}dx + \frac{\Theta_D}{4T}\right] + d^2 \qquad (2)$$

($\Theta_{E,ii}$ is the Einstein temperature; ii refers to the direction or plane of the vibration; *m* is the mean atomic mass; $k_B$ is the Boltzmann constant; $\hbar$ is the reduced Planck constant and $\Theta_D$ *is* the Debye temperature) result in $\Theta_{E,11}$ = 93 K, $\Theta_{E,22}$ = 67 K and $\Theta_D$ = 276 K for $Ba_{4.9}Sr_{3.1}Cu_{5.3}Ge_{40.7}$ and $\Theta_{E,11}$ = 68 K, $\Theta_{E,22}$ = 76 K and $\Theta_D$ = 268 K for $Sr_8Cu_{5.3}Ge_{40.7}$.

The linear thermal expansion coefficient, i.e.,

$$\alpha_a = \frac{1}{a_{300K}}\left(\frac{\partial a_T}{\partial T}\right)_p \quad , \tag{3}$$

where $a_{300K}$ is the lattice parameter at 300 K and $a_T$ is the lattice parameter at temperature T, is $\alpha_a = 12.65\times10^{-6}$ $K^{-1}$ ($Ba_{4.9}Sr_{3.1}Cu_{5.3}Ge_{40.7}$) and $\alpha_a = 13.25\times10^{-6}$ $K^{-1}$ ($Sr_8Cu_{5.3}Ge_{40.7}$).

The speed of sound ($\nu$) can be estimated from

$$\upsilon \approx \frac{2\pi\Theta_D k_B}{h}/\left(6\pi^2 n\right)^{1/3} \quad , \tag{4}$$

where n is the number of atoms per unit volume [6]. The calculated speed of sound according to equation (4) is 2500 m/s for $Sr_8Cu_{5.3}Ge_{40.7}$ and 2600 m/s for $Ba_{4.9}Sr_{3.1}Cu_{5.3}Ge_{40.7}$. These values range in the same order of magnitude as reported by Qiu et al. [6] for $Sr_8Zn_8Ge_{38}$ (2600 m/s) or $Sr_8Ga_{16}Ge_{30}$ (2800 m/s).

### *3.3 Phase relations at 700°C*

Information on the pertinent binary phase diagrams Sr-Ge and Cu-Ge is taken from Massalski [19]. Crystallographic data on the solid phases of the binary systems Sr-Ge and Cu-Ge and the ternary system Sr-Cu-Ge are summarized in Table I.

The phase equilibria in the Ge-rich part of the Sr-Cu-Ge system have been studied below the decomposition temperature of the clathrate type I at 700°C. Alloys in as cast state and after annealing are characterized by XPD and EPMA measurements. Only two three-phase equilibria exist in the region investigated (see figure 5). (Ge) is in equilibrium with $SrGe_2$ and $SrCu_{2-x}Ge_{2+x}$ ($\tau_1$) and with $\tau_1$ and liquid. $\tau_1$ crystallizes in the $ThCr_2Si_2$ structure type (XPD) at this temperature and its homogeneity range extends to higher Ge-contents of 49 at.% Ge (x=0.4; EPMA). The results of EPMA measurements and XPD are summarized in Table II (i and j).

### *3.4 Physical properties*

Sentence about getting almost single phase piece for phys. Prop.?

Temperature dependent electrical resistivity, ρ, studies performed for $Sr_8Cu_{5.3}Ge_{40.7}$ from 2.0

K to room temperature, revealed a simple metallic behavior as deduced from the constant slope of ρ(T) above 80 K (see figure 6). At lower temperatures, a combination of electron scattering on thermally activated acoustic phonons and scattering on low lying optical phonon branches allow to fairly well describe ρ(T), i.e.

$$\rho_{el}(T) = \rho_0 + 4A\cdot\left(\frac{T}{\Theta_D}\right)^5 \int_0^{\Theta_D/T}\left[\frac{x^5}{(e^x-1)(1-e^{-x})}dx\right] + B\cdot\left(\Theta_{E,11} - \Theta_{E,22} + T\cdot\ln\left(\frac{e^{\frac{\Theta_{E,22}}{T}}-1}{e^{\frac{\Theta_{E,11}}{T}}-1}\right)\right). \quad (5)$$

The first two terms refer to the conventional Bloch Grüneisen law, while the far right terms of Eqn. (5) represent the temperature dependence of ρ(T) originated from scattering of conduction electrons on optical phonons [35]. Employing the Einstein temperatures as derived above ($\Theta_{E,11}$ = 68 K and $\Theta_{E,22}$ = 76 K), reveals a reasonable fit to the data for a Debye temperature $\Theta_D$ = 193 K, a residual resistivity $\rho_0$ = 407 μΩcm and two material dependent parameters A = 75.3 μΩcm and B = 11.8 μΩcm/K. The Debye temperature from this procedure is smaller than that extracted from the ADP data. The resulting residual resistivity ratio (RRR) is 2.12. The rather high overall resistivity might be explained from a reduced charge carrier density due to the closeness of the material to a metal-insulator transition.

*3.4.2 Heat Capacity*

Specific heat measurements of $Sr_8Cu_{5.3}Ge_{40.7}$, taken from 2 K to 200 K, are plotted as $C_p/T$ vs $T$ in figure 7a. A standard low temperature fit $Cp(T) = \gamma T + \beta T^3$ was applied, resulting in a Debye temperature $\Theta_D^{LT}$ = 273 K that is close to the value obtained from the ADP calculations. The relatively large Sommerfeld value, γ = 24 mJ/molK$^2$, can be attributed to an electronic density of states at the Fermi energy of about 10 states/eV/f.u. if the electron-phonon enhancement factor is small. The lattice dynamics of $Sr_8Cu_{5.3}Ge_{40.7}$ can be derived, employing a model proposed by Junod et al. [36]. A comparison between a simple Debye model (red dashed line) and the Junod model (black dashed line) are presented in figure 7b. The heat capacity data obtained apparently deviate from the simple Debye scenario. As a consequence, an inclusion of Einstein modes with certain finite spectral widths is required to arrive at a proper description of $C_p(T)$. Such a fit renders a Debye temperature $\Theta_D$ = 258 K and two Einstein modes, $\Theta_{E,11}$

= 42 K and $\Theta_{E,22}$ = 75 K with a width of 1 K and 3 K, respectively. The resulting spectral function F(ω) is shown as a blue solid line in figure 7b, refering to the right axis. This analysis clarifies the contribution and backs the importance of Einstein modes in the present clathrate. A more detailed explanation of the models used can be found e.g., in Ref. [11].

*Conclusions*

In the Sr-Cu-Ge system, the clathrate type-I forms in a very small temperature range with a very limited solubility close to the Zintl limit $Sr_8Cu_{5.3}Ge_{40.7}$ and decomposes at about 730 °C eutectoidally into (Ge), $SrGe_2$ and $\tau_1$-$SrCu_{2-x}Ge_{2+x}$. At 700°C, a three phase equilibrium exists between (Ge), $SrGe_2$ and $\tau_1$. The ternary compound $\tau_1$-$SrCu_{2-x}Ge_{2+x}$ exhibit a homogeneity range in the Ge rich part to x=0.4 (a = 0.42850(4), c = 1.0370(1) nm; $ThCr_2Si_2$ type).

Lattice parameters of the metallic clathrate type-I solid solution $Ba_{8-x}Sr_xCu_yGe_{46-y}$ ($5.2 \le y \le 5.4$) decrease linearly with increasing Sr content and the maximum solubility of strontium is about 10 at%. The clathrate type-I crystal structure has been proven by X-ray single crystal diffraction on 2 single crystals with composition $Sr_8Cu_{5.3}Ge_{40.7}$ (a(Sr) = 1.06368(2) nm) and $Ba_{4.9}Sr_{3.1}Cu_{5.3}Ge_{40.7}$ (a(mix) = 1.06748(2) nm) measured at 300, 200 and 100 K. For the pure Sr compound an off-center model has been applied for the Sr atom inside the larger cage whereas in case of the mixed compound the guests take almost on-center positions in the cages. From the temperature dependency of the lattice parameter and the atomic displacement parameters, the thermal expansion coefficient ($\alpha$(Sr) = $13.25\cdot10^{-6}$ and $\alpha$(mix)= $12.65\cdot10^{-6}$ K), the Debye ($\Theta_D$(Sr)=268 and $\Theta_D$(mix)=276 K) and Einstein temperatures ($\Theta_{E,22}$(Sr)= 76, $\Theta_{E,11}$(Sr)= 68 and $\Theta_{E,22}$(mix)= 67, $\Theta_{E,11}$(mix)= 93 K) and the speed of sound (ν(Sr)=2.5 ν(mix)=2.6 km/s) have been determined. A detailed analysis of the heat capacity data of $Sr_8Cu_{5.3}Ge_{40.7}$ corroborates fairly well these Debye and Einstein temperatures. From these investigations, one might expect a very low and glass-like thermal conductivity as it has been reported for other Sr or Eu based clathrates type-I (see e.g., Ref. [37]).

*References*

1) H. Anno, M. Hokazono, H. Takakura, K. Matsubara, International Conference on Thermoelectrics (2005), 0-7803-9552-2/05

2) H. Anno, K. Suzuki, K. Koga, International Conference on Thermoelectrics (2007), 978-1-4244-2263-0/08

3) C. Candolfi, U. Aydemir, M. Baitinger, N. Oeschler, F. Steglich, Yu . Grin, JOURNAL OF APPLIED PHYSICS 111, 043706 (2012)

4) Zeiringer I, Chen MX, Bednar I, Royanian E, Bauer E, Podloucky R, Grytsiv A, Rogl P, Effenberger H.: Acta. Mater. 2011; 59: 2368-2384

5) B. Eisenmann, H. Schäfer, R. Zagler, Journal of the Less-Common Metals 118 (1986) 43-55

6) Qiu Liyan; Swainson, I.P.; Nolas, G.S.; White, M.A., Physical Review B 70 (2004), 035208-1-035208-8

7) D.C. Li, L. Fang, S.K. Deng, K.Y. Kang, W.H. Wei, H.B. Ruan, Phys. Status Solidi B 249, No. 7, 1423–1430 (2012)

8) Kengo Kishimoto, Naoya Ikeda, Koji Akai, Tsuyoshi Koyanagi, Applied Physics Express 1 (2008) 031201

9) H. Anno, H. Fukushima, K. Koga, K. Okiza, K. Matsubara, International Conference on Thermoelectrics (2006), 1-4244-0811-3/06

10) J.H. Roudebush, E.S. Toberer, H. Hope, G.F. Snyder, S.M. Kauzlarich, Journal of Solid State Chemistry 184 (2011) 1176–1185

11) Melnychenko-Koblyuk N, Grytsiv A, Berger St, Kaldarar H, Michor H, Rohrbacher F, Royanian E, Bauer E, Rogl P, Schmid H, Giester G, J. Phys. Condens. Matter 19 (2007) 046203

12) Zeiringer I, Grytsiv A, Broz P, Rogl P, J. of Solid State Chemistry 2012; 196; 125-131

13) Oscail: Patrick McArdle, Karen Gilligan, Desmond Cunningham, Rex Dark and Mary Mahon;  CrystEngComm, 2004, 6, 303 – 309

14) SHELX - G.M. Sheldrick, Acta Cryst. A, 2008, 64, 112-122

15) Gelato ML, Parthe E. J Appl Cryst 1987;20:139

16) Yuegang Zhang, Peter L. Lee, George. S. Nolas, Angus P. Wilkinson; Applied Physics Letters 80 (2002), 16, 2931

17) S. Paschen, S. Budnyk, U.Köhler, Yu. Prots, K. Hiebl, F. Steglich, Yu. Grin; Physica 383 (2006) 89-92

18) Ya Mudryk, P Rogl, C Paul, S Berger, E Bauer, G Hilscher, C Godart, H Noel; J. Phys.: Condens. Matter 14 (2002) 7991–8004

19) Binary Alloy Phase Diagrams, 2nd Edition, Ed. T.B. Massalski, ASM International, Materials Park, Ohio (1990) 1

20) Betz A, Schäfer H, Weiss A, Wulf R; Zeitschrift für Naturforschung B (1968) 23, 878

21) Lenz J, Schubert K; Zeitschrift für Metallkunde (1971) 62, 810

22) Schubert K, Brandauer G, Zeitschrift für Metallkunde (1952) 43, 262

23) Rieger, W.;Parthe, E.; Monatshefte fuer Chemie und verwandte Teile anderer Wissenschaften (1969), 100, 444-454

24) Yang Li, Ji Chi, Weiping Gou, Sameer Khandekar and Joseph H Ross Jr; J. Phys.: Condens. Matter 15 (2003) 5535–5542

25) Schäfer MC, Yamasaki Y, Fritsch V, Bobev S; Z. Anorg. Allg. Chemie (2012) 638 (7-8); 1204-1212

26) C. Kranenberg, A. Mewis, Z. Anorg. Allg. Chem. 2003, 629, 1023-1026

27) Sheldon E.A., King A.J., Acta Crystallogr. (1953) 6, 100

28) Betz A., Schäfer H.,  Weiss A., Z. Naturforsch. B (1967) 22, 103

29) Eisenmann B., Schäfer H., Turban K., Z. Naturforsch. B (1975) 30, 677-680

30) Lenz J., Schubert K., Z. Metallkd. (1971) 62, 810-816

31) Schubert K., Breimer H., Burkhardt W., Günzel E., Haufler R., Lukas H.L., Vetter H., Wegst J., Wilkens M., Naturwissenschaften (1957) 44, 229-230

32) Eisenmann B., Schäfer H., Turban K., Z. Naturforsch. B (1974) 29, 464-468

33) Rieger W., Parthé E., Monatsh. Chem. (1969) 100, 439-443

34) Dörrscheidt W., Niess N., Schäfer H., Z. Naturforsch. B (1976) 31, 890-891

35) H.-L. Engquist, Physical Review B, 21 p. 2067 (1980)

36) A. Junod, D. Bichsel, and J. Muller, Helv. Phys. Acta 52, 580 (1979).; A. Junod, Physical Review B, 27 p. 1568 (1983)

37) F. Bridges and L. Downward, Physical Review B 70 (2004) 140102

Table I: Crystallographic data on the solid phases of the binary systems Sr-Ge and Cu-Ge and the ternary system Sr-Cu-Ge

| **Phase** | **Space group** | **Structure type** | **Lattice parameter [nm]** | | | **Ref.** |
|---|---|---|---|---|---|---|
| | | | **a** | **b** | **c** | |
| (Ge) | $Fd\bar{3}m$ | $C_{diam}$ | 0.56574 | - | - | [19] |
| (Sr) hT | $Im\bar{3}m$ | W | 0.485(1) | - | - | [27] |
| (Sr) rT | $Fm\bar{3}m$ | Cu | 0.60849(5) | - | - | [27] |
| (Cu) | $Fm\bar{3}m$ | Cu | 0.36416 | - | - | [19] |
| $SrGe_2$ | *Pnma* | $BaSi_2$ | 0.874 | 0.665 | 1.124 | [20] |
| SrGe | *Cmcm* | TlI | 0.486(1) | 1.140(1) | 0.419(1) | [28] |
| $SrGe_{0.76}$ rT | *Immm* | SrSi | 0.484 | 1.338 | 1.852 | [32] |
| $Sr_2Ge$ | *Pnma* | $PbCl_2$ | 0.813(2) | 0.520(6) | 0.958(2) | [29] |
| $Cu_{2.5}Ge$ hT | $Fm\bar{3}m$ | $BiF_3$ | 0.5906 | - | - | [30] |
| $Cu_3Ge$ hT | $P6_3/mmc$ | $IrAl_3$ | 0.4169 | - | 0.7499 | [21] |
| $Cu_3Ge$ rT | *Pmmn* | $Cu_3Ti$ | 0.455 | 0.529 | 0.420 | [31] |
| $Cu_{0.85}Ge_{0.15}$ | $P6_3/mmc$ | Mg | 0.2612 | - | 0.4232 | [22] |
| $\kappa_I$-$Sr_8Cu_xGe_{46-x}$ | $Pm\bar{3}n$ | $K_4Ge_{23-x}$ | 1.06307(2) | - | - | x=5.3* |
| $SrCu_{0.67}Ge_{1.33}$ | *P6/mmm* | $AlB_2$ | 0.4230 | - | 0.4619 | [33] |
| $SrCu_9Ge_4$ | *I4/mcm* | $CeNi_{8.5}Si_{4.5}$ | 0.8273(2) | - | 1.1909(5) | [26] |
| $\tau_1$-$SrCu_{2-x}Ge_{2+x}$ | *I4/mmm* | $ThCr_2Si_2$ | 0.4270(6)<br>0.4256(2)<br>0.428(1)<br>0.42748(3)<br>0.42850(4) | -<br>-<br>-<br>- | 1.0258(10)<br>1.0205(8)<br>1.031(2)<br>1.03751(9)<br>1.0370(1) | x=0 [23]<br>x=0 [25]<br>x=0 [34]<br>x=0.5*<br>x=0.4* |

* This work

Table II: EPMA and X-ray phase analysis data corresponding to the microstructures of the alloys shown in figure 1

| **Alloy** | **X-ray phase analysis** | **Structure type** | **Lattice parameter [nm]** | | | **Composition EPMA (at.%)** | | | |
|---|---|---|---|---|---|---|---|---|---|
| | | | **a** | **b** | **c** | **Sr** | **Ba Si Sn** | **Cu** | **Ge** |
| $Sr_8Cu_{5.3}Ge_{40.7}$ | (Ge) | $C_{diam}$ | 0.56579(5) | - | - | 0.0 | - | 1.6 | 98.4 |
| a) and b) | $\kappa_I$ | $K_4Ge_{23-x}$ | 1.06307(2) | - | - | 14.3 | - | 9.8 | 75.9 |
| 600°C | $SrGe_2$ | $BaSi_2$ | - | - | - | 31.7 | - | 1.2 | 67.1 |
| | $\tau_1$ | $ThCr_2Si_2$ | 0.42871(8) | - | 1.0370(1) | 19.3 | - | 31.4 | 49.3 |
| $Sr_8Cu_{5.3}Ge_{40.7}$ | (Ge) | $C_{diam}$ | 0.56569(5) | - | - | 0.0 | - | 1.6 | 98.4 |
| c) | $\kappa_I$ | $K_4Ge_{23-x}$ | 1.06288(4) | - | - | 14.2 | - | 9.5 | 76.3 |
| as cast | $SrGe_2$ | $BaSi_2$ | - | - | - | 31.1 | - | 1.5 | 67.4 |
| | $\tau_1$ | - | - | - | - | 19.2 | - | 20.9 | 59.9 |
| $Sr_8Cu_5Ge_{41}Si_5$ | (GeSi) | $C_{diam}$ | - | - | - | - | 6-30 | ~1.7 | 94-70 |
| d) as cast | $\tau_2$ | - | - | - | - | 23.3 | 2.6 | 20.9 | 53.2 |
| | $SrGe_2$ | $BaSi_2$ | - | - | - | 30.8 | - | 1.9 | 67.3 |
| $Sr_8Cu_5Ge_{44}Sn_2$ | (Ge) | $C_{diam}$ | - | - | - | - | 1.2 | 1.7 | 97.1 |
| e) as cast | $\tau_3$ | - | - | - | - | 21.9 | 21.8 | 0.9 | 55.4 |
| | $SrGe_2$ | $BaSi_2$ | - | - | - | 32.1 | 1.3 | 1.7 | 64.9 |
| $Ba_{0.5}Sr_{7.5}Cu_{5.3}Ge_{40.7}$ | $\kappa_I$ | $K_4Ge_{23-x}$ | 1.06277(3) | - | - | 14.6 | - | 9.4 | 75.9 |
| f) as cast | $\kappa_{I\text{-}Sr,Ba}$ | $K_4Ge_{23-x}$ | 1.0648(1) | - | - | 9.9 | 4.3 | 9.5 | 76.3 |
| | $SrGe_2$ | $BaSi_2$ | - | - | - | 31.6 | - | 1.4 | 67.0 |
| $Ba_{0.5}Sr_{7.5}Cu_{5.3}Ge_{40.7}$ | $\kappa_I$ | $K_4Ge_{23-x}$ | 1.06315(3) | - | - | 14.6 | - | 9.6 | 75.9 |
| g) 760°C | $\kappa_{I\text{-}Sr,Ba}$ | $K_4Ge_{23-x}$ | - | - | - | 8.8 | 6.0 | 9.7 | 75.6 |
| | $SrGe_2$ | $BaSi_2$ | - | - | - | 32.0 | - | 1.6 | 66.4 |
| $Ba_{0.5}Sr_{7.5}Cu_{5.3}Ge_{40.7}$ | $\kappa_I$ | $K_4Ge_{23-x}$ | - | - | - | 14.4 | - | 9.9 | 75.7 |
| h) 700°C | $\kappa_{I\text{-}Sr,Ba}$ | $K_4Ge_{23-x}$ | 1.06588(6) | - | - | 7.4 | 7.2 | 9.9 | 75.5 |
| | | | - | - | - | 9.6 | 4.9 | 9.8 | 75.7 |
| | (Ge) | $C_{diam}$ | 0.56574(3) | - | - | 0.1 | - | 1.3 | 98.6 |
| | $SrGe_2$ | $BaSi_2$ | - | - | - | 32.0 | - | 1.1 | 66.9 |
| | $\tau_1$ | $ThCr_2Si_2$ | - | - | - | 19.5 | - | 30.8 | 49.7 |
| $SrCuGe_3$ | (Ge) | $C_{diam}$ | 0.56579(3) | - | - | 0.0 | - | 0.5 | 99.5 |
| i) 700°C | $SrGe_2$ | $BaSi_2$ | - | - | - | 31.9 | - | 1.3 | 66.8 |
| | $\tau_1$ | $ThCr_2Si_2$ | 0.42850(4) | - | 1.0370(1) | 19.9 | - | 31.4 | 48.7 |
| $SrCu_4Ge_{12}$ | (Ge) | $C_{diam}$ | 0.56568(5) | - | - | 0.0 | - | 0.7 | 99.3 |
| j) 700°C | $\tau_1$ | $ThCr_2Si_2$ | 0.42700(5) | - | 1.0376(1) | 20.0 | - | 34.9 | 45.1 |
| | eutectic | - | - | - | - | 0.9 | - | 57.8 | 41.3 |

Table III: X-Ray single crystal data for $Ba_{4.9}Sr_{3.1}Cu_{5.3}Ge_{40.7}$ at 300 K standardized with the program *Structure Tidy* (MoKα-radiation; 2°≤2Θ≤70°; ω-scans, scan width 2°; 150 sec/frame; Anisotropic displacement parameters $U_{ij}$ in [Å$^2$])

| **Parameter/compound** | **$Ba_{4.9}Sr_{3.1}Cu_{5.3}Ge_{40.7}$** |
|---|---|
| Space Group | $Pm\bar{3}n$ |
| Formula from EPMA | $Ba_4Sr_4Cu_{5.3}Ge_{40.7}$ |
| Formula from refinement | $Ba_{4.86}Sr_{3.14}Cu_{5.36}Ge_{40.64}$ |
| *a* [nm] | 1.06748(2) |
| $\mu_{abs}$ [mm$^{-1}$] | 34.52 |
| *V* (nm$^3$) | 1.216 |
| $\rho_x$ (gcm$^{-3}$) | 5.7 |
| Reflections in refinement | 439≥4σ($F_o$) of 561 |
| Number of variables | 20 |
| $R_F = \Sigma\|F_0-F_c\|/\Sigma F_0$ | 0.022 |
| $R_{Int}$ | 0.057 |
| wR2 | 0.043 |
| GOF | 1.152 |
| Extinction (Zachariasen) | 0.00109(9) |
| Residual density e$^-$/Å$^3$; max; min | 1.539; -1.391 |
| Atom parameters | |
| AE1 in 2a (0,0,0); occ. | 0.195(4) Ba + 0.805 Sr |
| $U_{11}= U_{22}=U_{33}$ | 0.0125(3) |
| AE2 in 6c (1/4,0,1/2); occ. | 0.745(3) Ba + 0.255 Sr |
| $U_{11}$; $U_{22} = U_{33}$ | 0.0239(4); 0.0431(3) |
| M in 6d (1/4,1/2,0); occ. | 0.107(4) Ge + 0.893 Cu |
| $U_{11}$; $U_{22} = U_{33}$ | 0.0145(5); 0.0102(3) |
| Ge2 in 16i (*x*, *x*, *x*); occ. | 1.000(2) |
| *x* | 0.18296(3) |
| $U_{11}=U_{22}=U_{33}$ | 0.0096(1) |
| Ge3 in 24k (0,*y*,*z*); occ. | 0.999(2) |
| *y*, *z* | 0.11941(4), 0.31410(4) |
| $U_{11}$; $U_{22}$; $U_{33}$ | 0.0110(2); 0.0114(2); 0.0114(2) |
| Interatomic distances [nm]; Standard deviation < 0.0001 | |
| AE1 – 8 Ge2 0.3383 | –3 Ge3 0.2497 |
| –12Ge3 0.3587 | –1 AE1 0.3383 |
| AE2 – 8 Ge3 0.3562 | –3 Ge2 0.3906 |
| –4 M 0.3774 | –3 AE2 0.3972 |
| –8 Ge2 0.3972 | Ge3– 1 Ge3 0.2549 |
| M – 4 Ge3 0.2425 | –2 Ge2 0.2497 |
| –4 AE2 0.3774 | –1 M 0.2425 |
| –8 Ge2 0.3972 | –2 AE2 0.3561 |
| Ge2– 1 Ge2 0.2479 | –1 AE1 0.3587 |

Table IV: X-Ray single crystal data for $Sr_8Cu_{5.3}Ge_{40.7}$ at 300 K standardized with the program *Structure Tidy* (MoKα-radiation; 2°≤2Θ≤70°; ω-scans, scan width 2°; 150 sec/frame; Anisotropic displacement parameters $U_{ij}$ in [$Å^2$])

| **Parameter/compound** | **$Sr_8Cu_{5.3}Ge_{40.7}$** |
|---|---|
| Space Group | $Pm\bar{3}n$ |
| Formula from EPMA | $Sr_8Cu_{5.3}Ge_{40.7}$ |
| Formula from refinement | $Sr_8Cu_{5.36}Ge_{40.64}$ |
| *a* [nm] | 1.06368(2) |
| $\mu_{abs}$ [$mm^{-1}$] | 36.07 |
| *V* ($nm^3$) | 1.203 |
| $\rho_x$ ($gcm^{-3}$) | 5.5 |
| Reflections in refinement | 423≥4σ($F_o$) of 548 |
| Number of variables | 22 |
| $R_F = \Sigma\|F_0-F_c\|/\Sigma F_0$ | 0.020 |
| $R_{Int}$ | 0.029 |
| wR2 | 0.041 |
| GOF | 1.076 |
| Extinction (Zachariasen) | 0.0013(1) |
| Residual density $e^-/Å^3$; max; min | 1.029; -1.228 |
| Atom parameters | |
| Sr1 in 2a (0,0,0); occ. | 0.989(4) |
| $U_{11}=U_{22}=U_{33}$ | 0.0135(3) |
| Sr2 in 24k (0,*y*,*z*); occ. | 0.248(1) |
| *y*, *z* | 0.2405(6), 0.4771(7) |
| $U_{11}$; $U_{22}=U_{33}$ | 0.028(1); 0.043(1) |
| M in 6c (1/4,0,1/2); occ. | 0.106(3) Ge + 0.894 Cu |
| $U_{11}$; $U_{22}=U_{33}$ | 0.0155(4); 0.0108(2) |
| Ge2 in 16i (*x*, *x*, *x*); occ. | 1.003(2) |
| *x* | 0.1828(2) |
| $U_{11}=U_{22}=U_{33}$ | 0.0104(1) |
| Ge3 in 24k (0,*y*,*z*); occ. | 0.998(2) |
| *y*, *z* | 0.3148(4), 0.1187(5) |
| $U_{11}$; $U_{22}$; $U_{33}$ | 0.0119(1); 0.0115(1); 0.0122(1) |

Interatomic distances [nm]; Standard deviation < 0.0001

| | | | | | |
|---|---|---|---|---|---|
| Sr1 – | 8Ge2 | 0.3369 | M – | 4Ge3 | 0.2414 |
| | –12Ge3 | 0.3579 | | –4Sr2 | 0.3667 |
| Sr2 – | 2Ge3 | 0.3335 | | –8Sr2 | 0.3698 |
| | – 2Ge3 | 0.3541 | | –8Ge2 | 0.3959 |
| | – 2Ge3 | 0.3661 | | –4Sr2 | 0.4005 |
| | – 1M | 0.3668 | Ge2– | 1Ge2 | 0.2474 |
| | – 2M | 0.3698 | | –3Ge3 | 0.2494 |
| | – 2Ge3 | 0.3710 | | –Sr1 | 0.3369 |
| | – 2Ge2 | 0.3736 | | –3Sr2 | 0.3796 |
| | – 2Ge2 | 0.3865 | | –3Sr2 | 0.3865 |
| | – 1Ge3 | 0.3893 | | –3Ge2 | 0.3890 |
| | – 1M | 0.4005 | Ge3– | 1M | 0.2414 |
| | – 2Ge2 | 0.4102 | | –2Ge2 | 0.2494 |
| | – 2Ge3 | 0.4105 | | –1Ge3 | 0.2526 |
| | – 1Ge3 | 0.4370 | | –2Sr2 | 0.3335 |

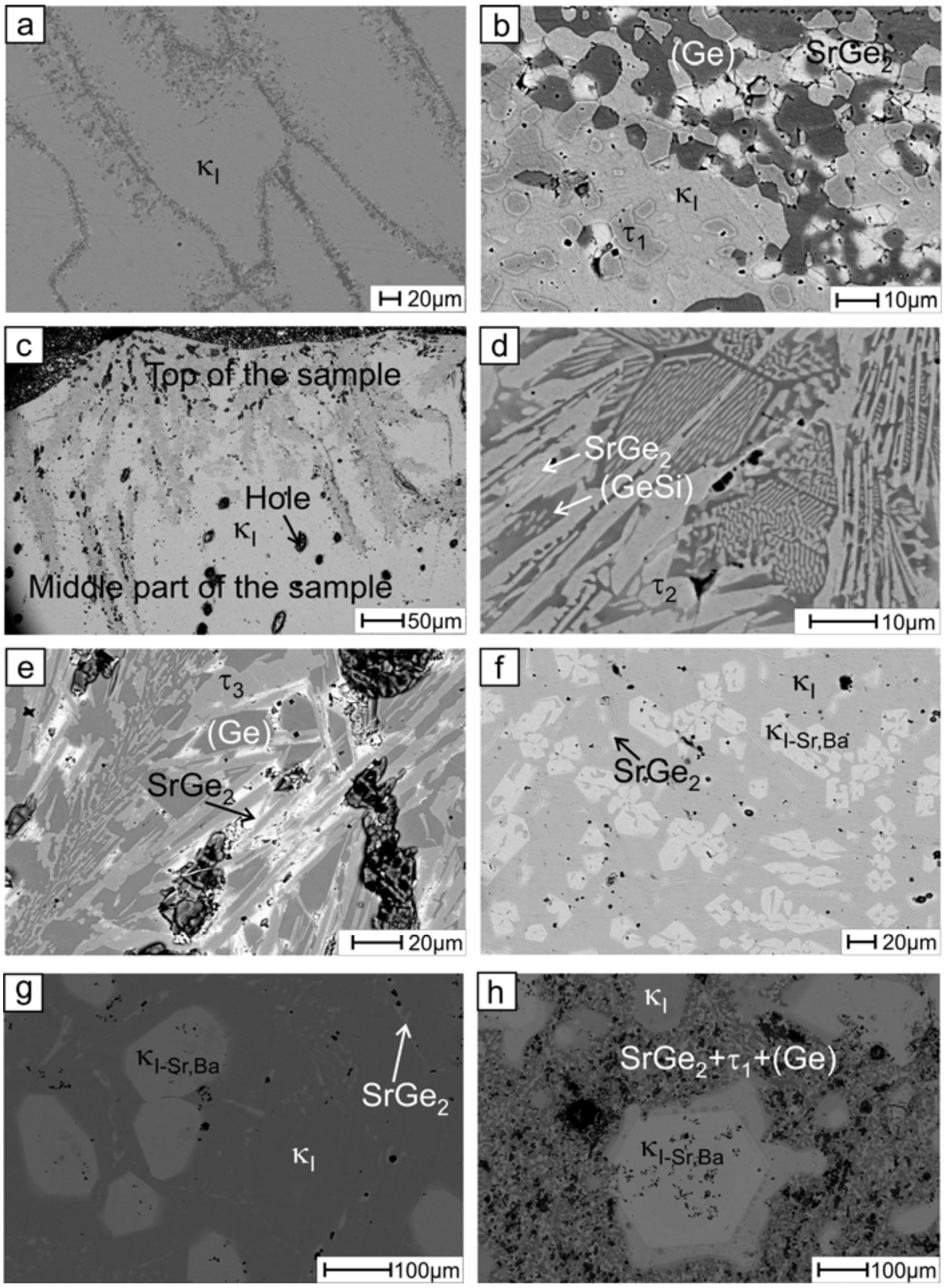

Fig. 1: EPMA images of alloys with nominal composition $Sr_8Cu_{5.3}Ge_{40.7}$ annealed at 600°C (a), partial decomposition (b) and in as cast state (c), of $Sr_8Cu_5Ge_{41}Si_5$ in as cast state (d), of $Sr_8Cu_5Ge_{44}Sn_2$ in as cast state (e) and of $Ba_{0.5}Sr_{7.5}Cu_{5.3}Ge_{40.7}$ in as cast state (f), annealed at 760°C (g) and 700°C (h); composition of all phases from EPMA measurements can be found in Table II

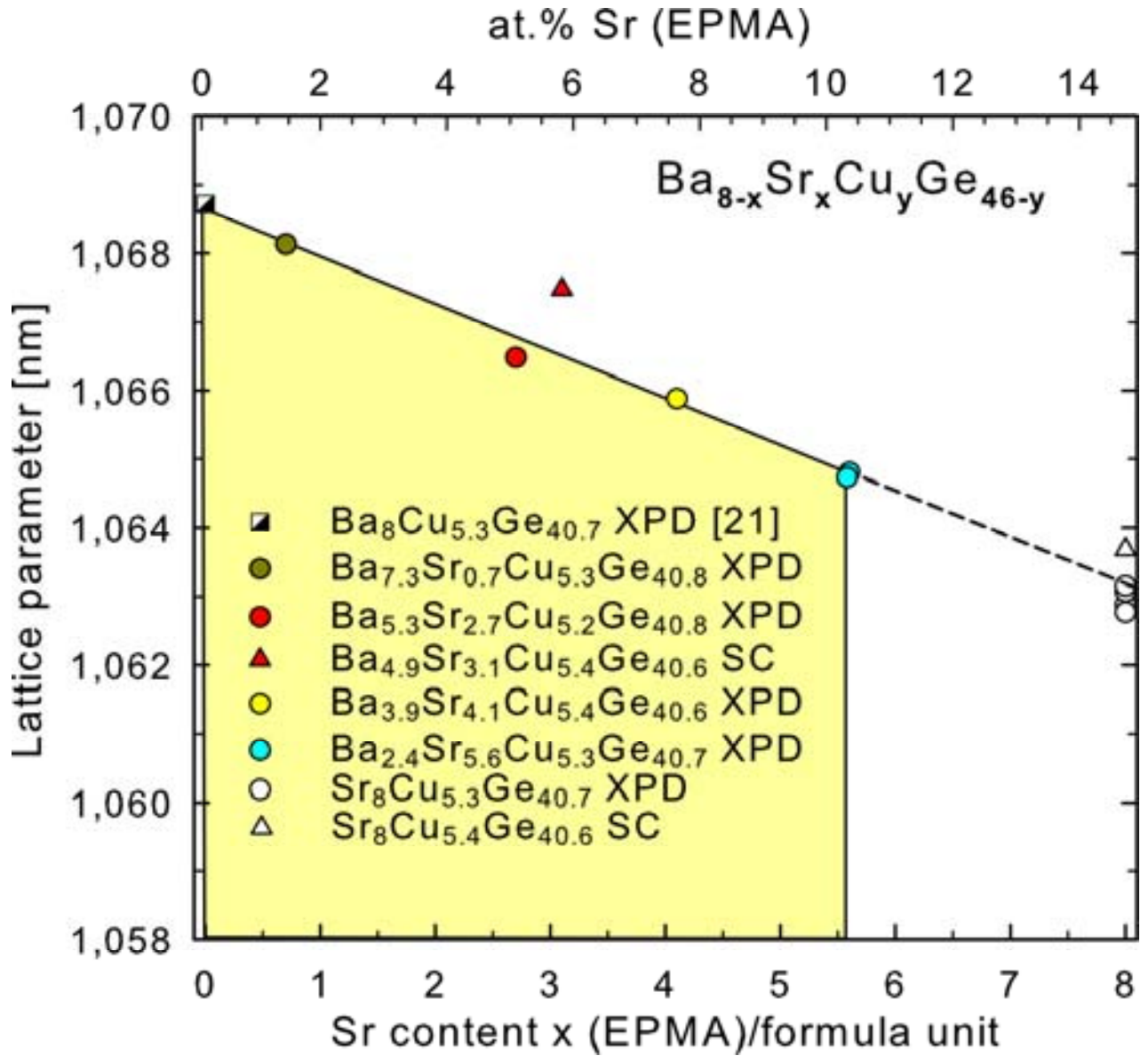

Fig. 2: Dependency of the lattice parameter with increasing Sr-content for the clathrate type-I solid solution $Ba_{8-x}Sr_xCu_yGe_{46-y}$ with $5.2 \le y \le 5.4$; the lattice parameters were obtained either from single crystal X-ray diffraction (SC) or from X-ray powder diffraction (XPD) using Ge as internal standard; the yellow area indicates the solubility limit of Sr determined by EPMA in as cast stat

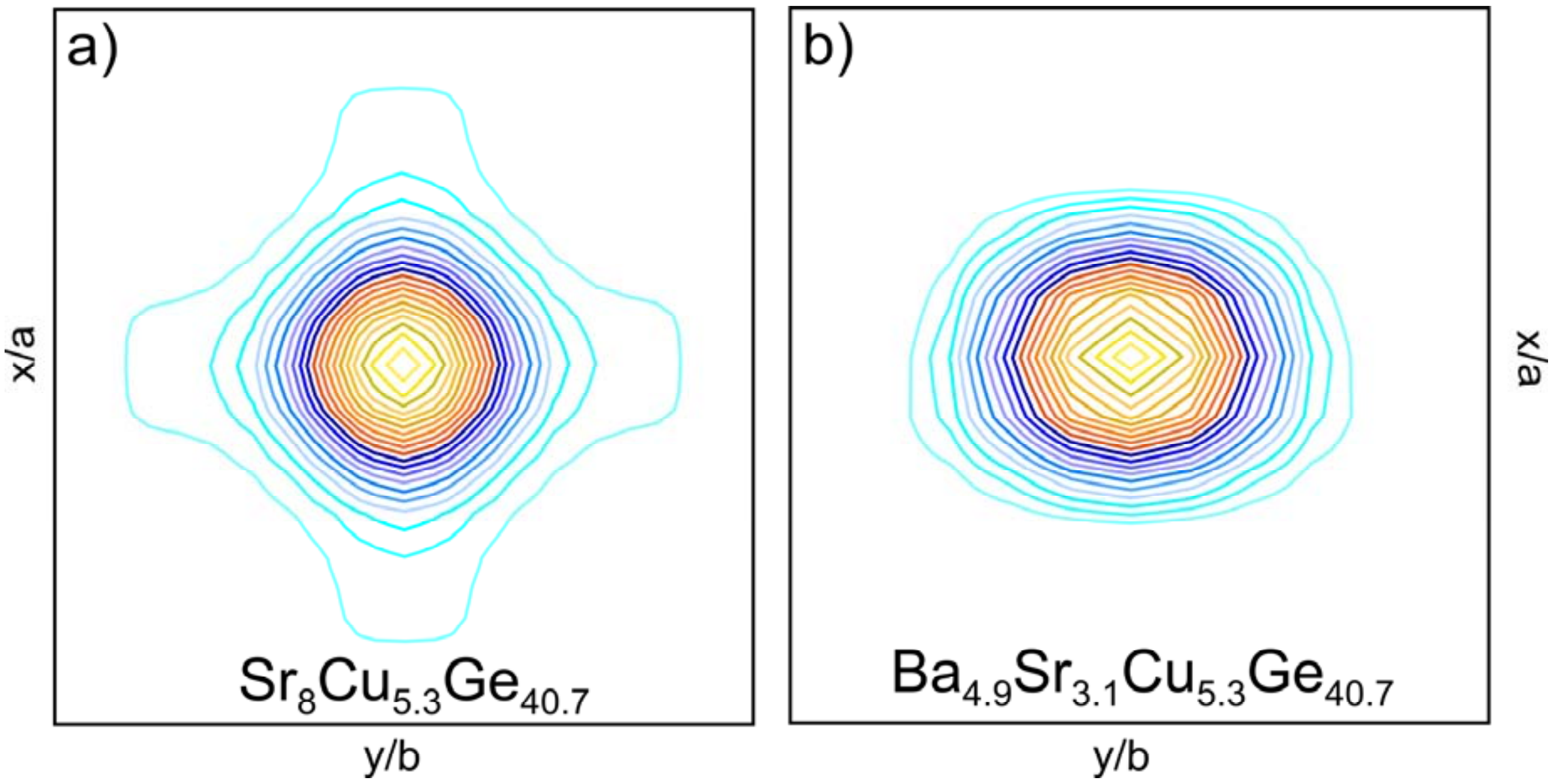

Fig. 3: Contour plot of the electron density at 300 K of the atoms in 6c-site for (a) $Sr_8Cu_{5.3}Ge_{40.7}$ and (b) $Ba_{4.9}Sr_{3.1}Cu_{5.3}Ge_{40.7}$ from difference Fourier synthesis; negative electron density has been omitted

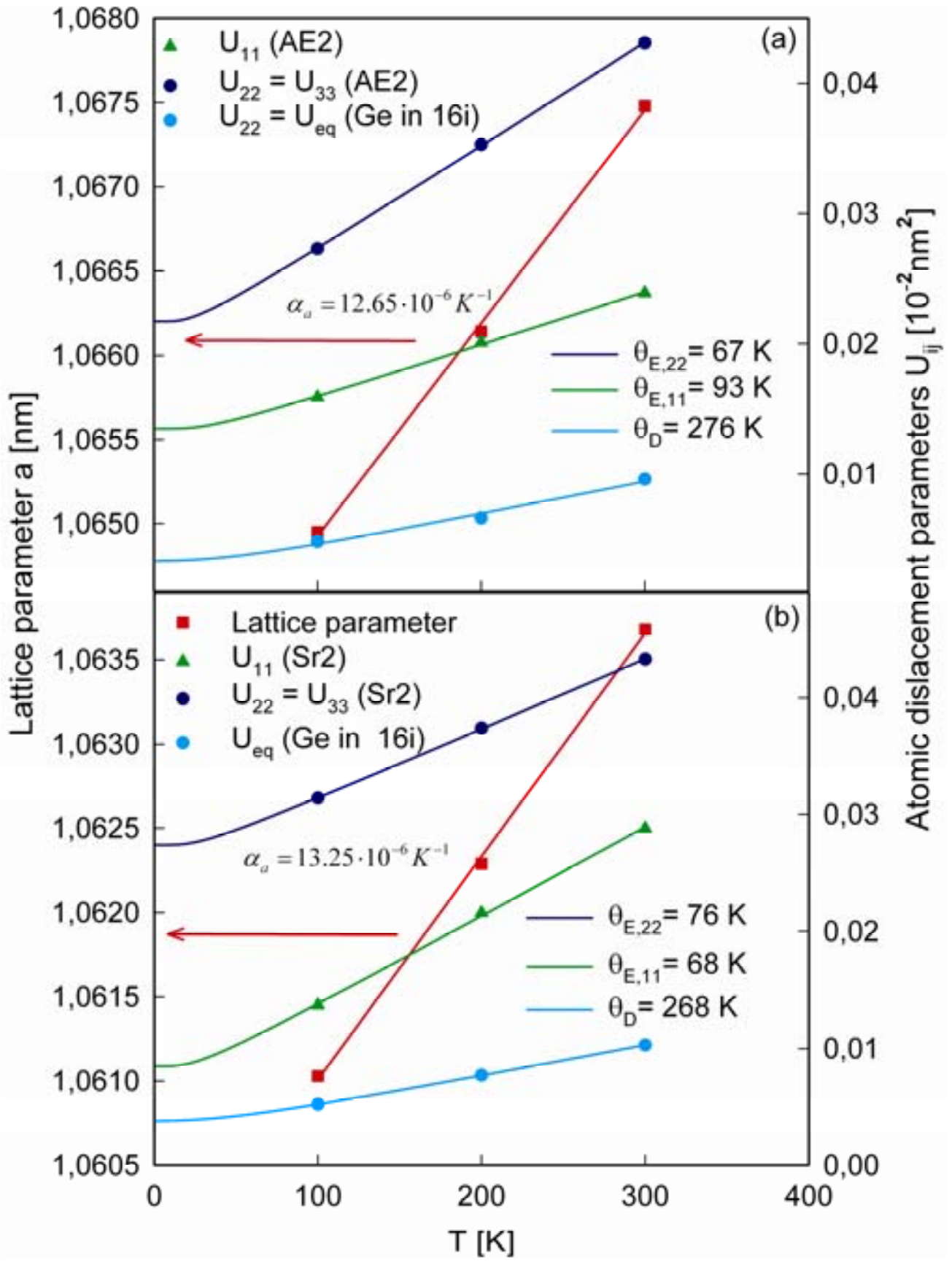

Fig. 4: Temperature dependent lattice parameter and atomic displacement parameters of $Ba_{4.9}Sr_{3.1}Cu_{5.3}Ge_{40.7}$ (a) and $Sr_8Cu_{5.3}Ge_{40.7}$ (b)

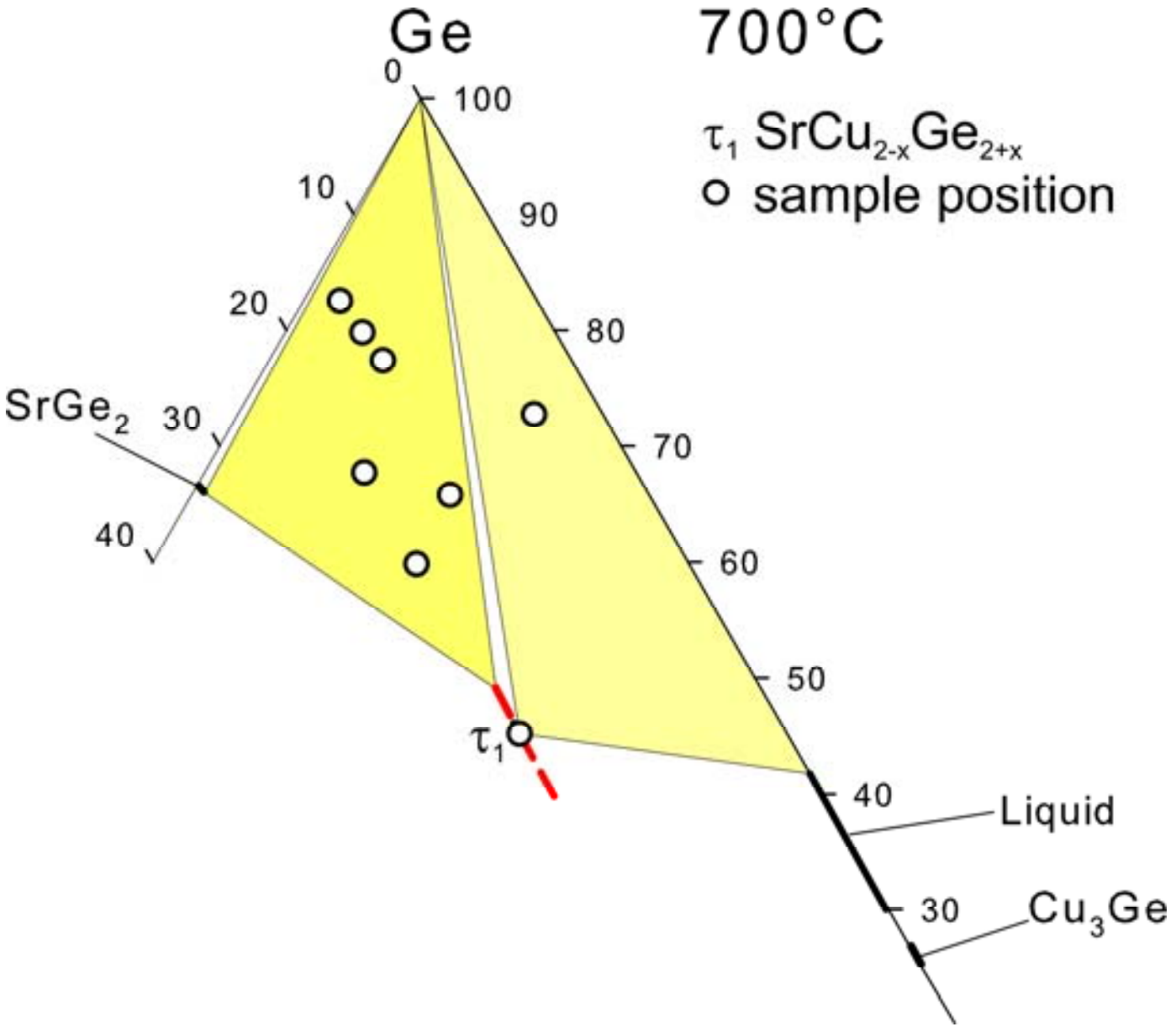

Fig. 5: Partial isothermal section of the Sr-Cu-Ge system at 700°C in the Ge-rich corner.

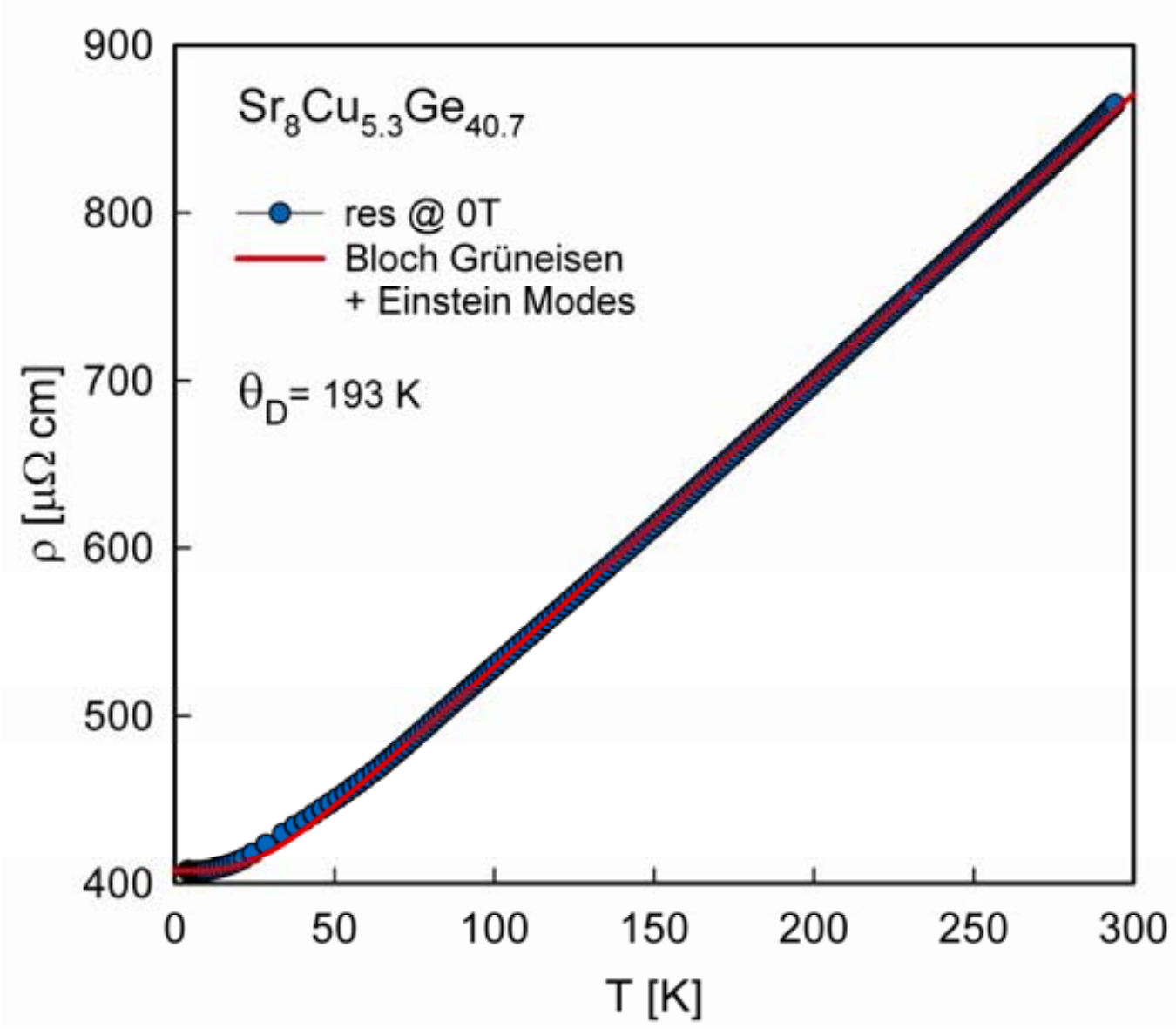

Fig. 6: Temperature dependent electrical resistivity, ρ, of $SrCu_{5.3}Ge_{40.7}$; the solid line corresponds to a least squares fit as explained in the text.

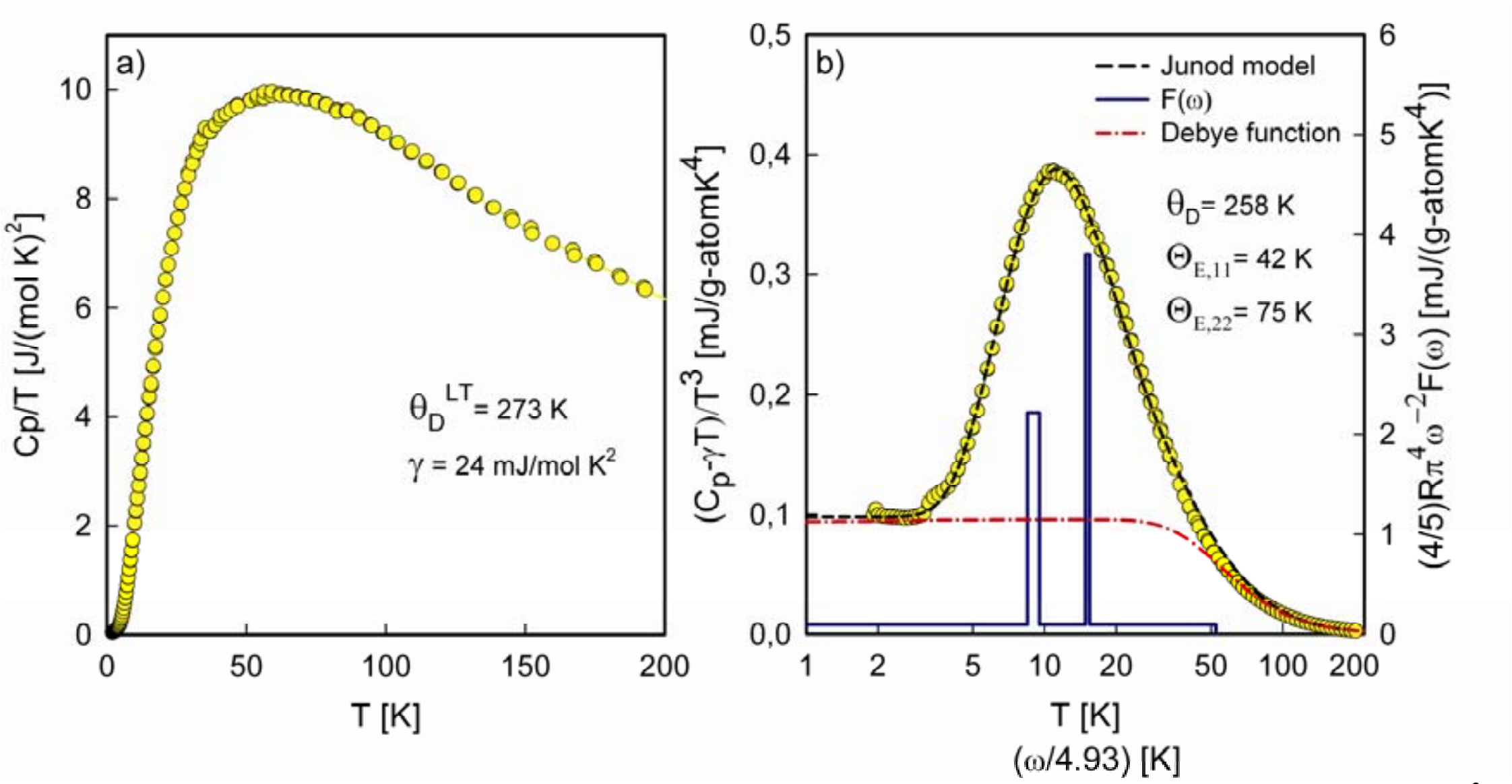

Fig. 7: Temperature dependent specific heat, $C_p$, of $Sr_8Cu_{5.3}Ge_{40.7}$ plotted as $C_p/T$ vs T (a), and $(Cp-\gamma T)/T^3$ vs lnT (b); the dashed line corresponds to the Junod fit as explained in the text. The spectral function, F(ω), plotted as $(4/5)R\pi^4\omega^{-2}F(\omega)$ vs. ω/4.93, is shown as a solid line (referring to the right axis); the dashed dotted line is the calculated Debye function

*Supplementary information*

Table I: X-Ray single crystal data for $Ba_{4.9}Sr_{3.1}Cu_{5.3}Ge_{40.7}$ at 200 K and 100 K standardized with the program *Structure Tidy* (MoKα-radiation; 2°≤2Θ≤70°; ω-scans, scan width 2°; 150 sec/frame; Anisotropic displacement parameters in [Å$^2$])

| **Parameter/compound** | **200 K** | **100 K** |
|---|---|---|
| Space Group | $Pm\bar{3}n$ | $Pm\bar{3}n$ |
| Formula from EPMA | $Ba_4Sr_4Cu_{5.3}Ge_{40.7}$ | $Ba_4Sr_4Cu_{5.3}Ge_{40.7}$ |
| Formula from refinement | $Ba_{4.99}Sr_{3.01}Cu_{5.29}Ge_{40.71}$ | $Ba_{4.92}Sr_{3.08}Cu_{5.30}Ge_{40.70}$ |
| $a$ [nm] | 1.06614(3) | 1.06495(3) |
| $\mu_{abs}$ [mm$^{-1}$] | 34.65 | 34.76 |
| $V$ (nm$^3$) | 1.211 | 1.207 |
| $\rho_x$ (gcm$^{-3}$) | 5.7 | 5.7 |
| Reflections in refinement | 460≥4σ($F_o$) of 559 | 464≥4σ($F_o$) of 555 |
| Number of variables | 19 | 19 |
| $R_F = \Sigma\|F_0-F_c\|/\Sigma F_0$ | 0.023 | 0.025 |
| $R_{Int}$ | 0.025 | 0.025 |
| wR2 | 0.044 | 0.055 |
| GOF | 1.177 | 1.145 |
| Extinction (Zachariasen) | 0.00065(7) | 0.0005(1) |
| Residual density e$^-$/Å$^3$; max; min | 1.69; -1.56 | 1.98; -2.43 |
| Atom parameters | | |
| AE1 in 2a(0,0,0); occ. | 0.182(4) Ba + 0.818 Sr | 0.212(5) Ba + 0.788 Sr |
| $U_{11}=U_{22}=U_{33}$ | 0.0086(3) | 0.0067(4) |
| AE2 6c (1/4,0,1/2); occ. | 0.772(3) Ba + 0.228 Sr | 0.749(4) Ba + 0.251 Sr |
| $U_{11}$; $U_{22}=U_{33}$ | 0.0201(4); 0.0353(3) | 0.0159(4); 0.0273(3) |
| M in 6d (1/4,1/2,0); occ. | 0.119(5) Ge + 0.881 Cu | 0.117(6) Ge + 0.883 Cu |
| $U_{11}$; $U_{22}=U_{33}$ | 0.0107(5); 0.0074(3) | 0.0076(5); 0.0050(3) |
| Ge2 in 16i($x$, $x$, $x$); occ. | 1.004(3) | 1.002(3) |
| $x$ | 0.18298(3) | 0.18297(3) |
| $U_{11}=U_{22}=U_{33}$ | 0.0066(1) | 0.0048(1) |
| Ge3 in 24k(0,$y$,$z$); occ. | 1.002(2) | 1.001(2) |
| $y$, $z$ | 0.11941(4), 0.31419(4) | 0.11943(4), 0.31421(5) |
| $U_{11}$; $U_{22}$; $U_{33}$ | 0.0079(2); 0.0084(2); 0.0083(2) | 0.0057(2); 0.0066(2); 0.0065(2) |
| Interatomic distances [nm]; Standard deviation < 0.0001 | | |
| | AE1 – 8Ge2 0.3379 | AE1 – 8Ge2 0.3375 |
| | –12Ge3 0.3583 | –12Ge3 0.3580 |
| | AE2 – 8Ge3 0.3557 | AE2 – 8Ge3 0.3553 |
| | – 4M 0.3769 | – 4M 0.3765 |
| | – 8Ge2 0.3967 | – 8Ge2 0.3963 |
| | – 4Ge3 0.4115 | – 4Ge3 0.4110 |
| | M – 4Ge3 0.2421 | M – 4Ge3 0.2418 |
| | –4AE2 0.3769 | –4AE2 0.3765 |
| | –8Ge2 0.3967 | –8Ge2 0.3963 |
| | –4Ge3 0.4409 | –4Ge3 0.4404 |
| | Ge2– 1Ge2 0.2475 | Ge2– 1Ge2 0.2473 |
| | –3Ge3 0.2494 | –3Ge3 0.2492 |
| | –1AE1 0.3379 | –1AE1 0.3375 |
| | –3Ge2 0.3902 | –3Ge2 0.3897 |
| | –3AE2 0.3967 | –3AE2 0.3963 |
| | –3M 0.3967 | –3M 0.3963 |
| | –3Ge3 0.3983 | –3Ge3 0.3978 |
| | Ge3– 1M 0.2421 | Ge3– 1M 0.2418 |
| | –2Ge2 0.2494 | –2Ge2 0.2492 |
| | –1Ge3 0.2546 | –1Ge3 0.2544 |
| | –2AE2 0.3557 | –2AE2 0.3553 |
| | –2AE1 0.3584 | –2AE1 0.3580 |

Table II: X-Ray single crystal data for $Sr_8Cu_{5.3}Ge_{40.7}$ at 200 K and 100 K standardized with the program *Structure Tidy* (MoKα-radiation; 2°≤2Θ≤70°; ω-scans, scan width 2°; 150 sec/frame; Anisotropic displacement parameters in [$Å^2$])

| **Parameter/compound** | **200 K** | **100 K** |
|---|---|---|
| Space Group | $Pm\bar{3}n$ | $Pm\bar{3}n$ |
| Formula from EPMA | $Sr_8Cu_{5.3}Ge_{40.7}$ | $Sr_8Cu_{5.3}Ge_{40.7}$ |
| Formula from refinement | $Sr_8Cu_{5.38}Ge_{40.62}$ | $Sr_8Cu_{5.36}Ge_{40.64}$ |
| $a$ [nm] | 1.06229 (2) | 1.06103(2) |
| $\mu_{abs}$ [$mm^{-1}$] | 36.21 | 36.34 |
| $V$ ($nm^3$) | 1.199 | 1.194 |
| $\rho_x$ ($gcm^{-3}$) | 5.5 | 5.5 |
| Reflections in refinement | 445≥4σ($F_o$) of 549 | 449≥4σ($F_o$) of 548 |
| Number of variables | 22 | 22 |
| $R_F = \Sigma\|F_0-F_c\|/\Sigma F_0$ | 0.023 | 0.022 |
| $R_{Int}$ | 0.026 | 0.026 |
| wR2 | 0.053 | 0.049 |
| GOF | 1.077 | 1.088 |
| Extinction (Zachariasen) | 0.0007(1) | 0.0005(1) |
| Residual density $e^-/Å^3$; max; min | 1.51; -0.98 | 1.76; -1.02 |
| Atom parameters | | |
| Sr1 in 2a(0,0,0); occ. | 1.0 | 1.0 |
| $U_{11}=U_{22}=U_{33}$ | 0.0108(2) | 0.0073(2) |
| Sr2 in 24k(0,$y$,$z$); occ. | 0.249(1) | 0.248(1) |
| $y$, $z$ | 0.2401(7), 0.4787(9) | 0.2395(6), 0.4801(8) |
| $U_{11}$; $U_{22}=U_{33}$ | 0.021(2); 0.039(2) | 0.026(1); 0.031(1) |
| M in 6c(1/4,0,1/2); occ. | 0.104(4) Ge + 0.896 Cu | 0.107(3) Ge + 0.893 Cu |
| $U_{11}$; $U_{22}=U_{33}$ | 0.0120(4); 0.0078(2) | 0.0083(4); 0.0051(3) |
| Ge2 in 16i($x$, $x$, $x$); occ. | 1.002(3) | 1.02(2) |
| $x$ | 0.18287(2) | 0.18285(2) |
| $U_{11}=U_{22}=U_{33}$ | 0.0077(1) | 0.0051(2) |
| Ge3 in 24k(0,$y$,$z$); occ. | 1.02(2) | 1.002(2) |
| $y$, $z$ | 0.31488(3), 0.11873(3) | 0.31493(3), 0.11873(3) |
| $U_{11}$; $U_{22}$; $U_{33}$ | 0.0093(2); 0.0088(2); 0.0094(2) | 0.0062(1); 0.0061(1); 0.0065(1) |
| Interatomic distances [nm]; Standard deviation < 0.0001 | | |
| | Sr1 – 8Ge2 0.3365 | Sr1 – 8Ge2 0.3360 |
| | –12Ge3 0.3575 | –12Ge3 0.3571 |
| | Sr2 – 2Ge3 0.3336 | Sr2 – 2Ge3 0.3335 |
| | – 2Ge3 0.3544 | – 2Ge3 0.3549 |
| | – 2Ge3 0.3593 | – 2Ge3 0.3574 |
| | – 1M 0.3677 | – 1M 0.3679 |
| | – 2M 0.3689 | – 2M 0.3688 |
| | – 2Ge3 0.3701 | – 2Ge3 0.3695 |
| | – 2Ge2 0.3744 | – 2Ge2 0.3753 |
| | – 2Ge2 0.3868 | – 2Ge2 0.3872 |
| | – 1Ge3 0.3905 | – 1Ge3 0.3917 |
| | – 1M 0.3991 | – 1M 0.3979 |
| | – 2Ge2 0.4083 | – 2Ge2 0.4077 |

| | | | | | | |
|---|---|---|---|---|---|---|
| | | – 2Ge3 | 0.4098 | | – 2Ge3 | 0.4092 |
| | | – 1Ge3 | 0.4131 | | – 1Ge3 | 0.4112 |
| | M – | 4Ge3 | 0.2411 | M – | 4Ge3 | 0.2407 |
| | | –4Sr2 | 0.3677 | | –8Sr2 | 0.3679 |
| | | –8Sr2 | 0.3689 | | –4Sr2 | 0.3668 |
| | | –8Ge2 | 0.3954 | | –8Ge2 | 0.3949 |
| | | –4Sr2 | 0.3991 | | –4Sr2 | 0.3979 |
| | Ge2– | 1Ge2 | 0.2470 | Ge2– | 1Ge2 | 0.2468 |
| | | –3Ge3 | 0.2491 | | –3Ge3 | 0.2488 |
| | | –Sr1 | 0.3365 | | –Sr1 | 0.3360 |
| | | –3Sr2 | 0.3774 | | –3Sr2 | 0.3752 |
| | | –3Sr2 | 0.3868 | | –3Sr2 | 0.3872 |
| | | –3Ge2 | 0.3885 | | –3G2 | 0.3880 |
| | Ge3– | 1M | 0.2411 | | 1M | 0.2407 |
| | | –2Ge2 | 0.2491 | | –2G2 | 0.2488 |
| | | –1Ge3 | 0.2522 | | –1G3 | 0.2519 |
| | | –2Sr2 | 0.3336 | | –2Sr2 | 0.3335 |